%% file: sample-sigconf.tex
\documentclass[sigconf]{acmart}
\usepackage{booktabs} 
\usepackage{tabularray}

\usepackage{fancyhdr,graphicx,amsmath,amssymb}

\usepackage[ruled,vlined]{algorithm2e}

\usepackage{background}
\usepackage{xcolor}
\usepackage{hyperref}

\hypersetup{
  pdfborder={0 0 0}, 
}

\copyrightyear{2024}
\acmYear{2024}
\setcopyright{acmlicensed}\acmConference[SAC '24]{The 39th ACM/SIGAPP Symposium on Applied Computing}{April 8--12, 2024}{Avila, Spain}
\acmBooktitle{The 39th ACM/SIGAPP Symposium on Applied Computing (SAC '24), April 8--12, 2024, Avila, Spain}
\acmPrice{15.00}
\acmDOI{10.1145/3605098.3635995}
\acmISBN{979-8-4007-0243-3/24/04}

\backgroundsetup{
  scale=1,
  color=black,
  opacity=1,
  angle=0,
  position=current page.south,
  vshift=10pt,
  contents={\textbf{\textcolor{red}{© 2024. For personal use only. Published version under doi:  \href{https://doi.org/10.1145/3605098.3635995}{10.1145/3605098.3635995}. }}}
}

\begin{document}

\title{Routing Optimization Based on Distributed Intelligent Network Softwarization for the Internet of Things}

\author{Mohamed Ali Zormati}
\affiliation{%
  \institution{Heudiasyc UMR 7253, Sorbonne University Alliance, Université de Technologie de Compiègne (UTC)}
  \city{Compiègne} 
  \country{France}
}
\email{zormati@ieee.org}

\author{Hicham Lakhlef}
\affiliation{%
  \institution{Heudiasyc UMR 7253, Sorbonne University Alliance, Université de Technologie de Compiègne (UTC)}
  \city{Compiègne} 
  \country{France}
}
\email{hicham.lakhlef@hds.utc.fr}

\author{Sofiane Ouni}
\affiliation{%
  \institution{National Institute of Applied Sciences and Technology (INSAT), University of Carthage}
  \city{Tunis} 
  \country{Tunisia}
}
\email{sofiane.ouni@insat.rnu.tn}

\renewcommand{\shortauthors}{Zormati, Lakhlef, and Ouni}

\begin{abstract}
The Internet of Things (IoT) establishes connectivity between billions of heterogeneous devices that provide a variety of essential everyday services. The IoT faces several challenges, including energy efficiency and scalability, that require consideration of enabling technologies such as network softwarization. This technology is an appropriate solution for IoT, leveraging Software Defined Networking (SDN) and Network Function Virtualization (NFV) as two main techniques, especially when combined with Machine Learning (ML). Although many efforts have been made to optimize routing in softwarized IoT, the existing solutions do not take advantage of distributed intelligence. In this paper, we propose to optimize routing in softwarized IoT networks using Federated Deep Reinforcement Learning (FDRL), where distributed network softwarization and intelligence (i.e., FDRL) join forces to improve routing in constrained IoT networks. Our proposal introduces the combination of two novelties (i.e., distributed controller design and intelligent routing) to meet the IoT requirements (mainly performance and energy efficiency). The simulation results confirm the effectiveness of our proposal compared to the conventional counterparts.
\end{abstract}

\begin{CCSXML}
<ccs2012>
   <concept>
       <concept_id>10003033.10003068</concept_id>
       <concept_desc>Networks~Network algorithms</concept_desc>
       <concept_significance>500</concept_significance>
       </concept>
   <concept>
       <concept_id>10010147.10010257.10010258.10010261</concept_id>
       <concept_desc>Computing methodologies~Reinforcement learning</concept_desc>
       <concept_significance>300</concept_significance>
       </concept>
   <concept>
       <concept_id>10010520.10010521.10010537</concept_id>
       <concept_desc>Computer systems organization~Distributed architectures</concept_desc>
       <concept_significance>300</concept_significance>
       </concept>
   <concept>
       <concept_id>10010520.10010553</concept_id>
       <concept_desc>Computer systems organization~Embedded and cyber-physical systems</concept_desc>
       <concept_significance>300</concept_significance>
       </concept>
 </ccs2012>
\end{CCSXML}

\ccsdesc[500]{Networks~Network algorithms}
\ccsdesc[300]{Computing methodologies~Reinforcement learning}
\ccsdesc[300]{Computer systems organization~Distributed architectures}
\ccsdesc[300]{Computer systems organization~Embedded and cyber-physical systems}

\keywords{Internet of Things (IoT), network softwarization, Federated Deep Reinforcement Learning (FDRL), routing optimization}

\maketitle

\input{samplebody-conf}

\bibliographystyle{ACM-Reference-Format}
\bibliography{sample-bibliography} 

\end{document}

%% file: samplebody-conf.tex
\section{Introduction}
The Internet of Things (IoT) is a promising technology that enables connectivity of anything, anywhere, at any time \cite{xie22}. The IoT ecosystem now includes billions of devices that affect every aspect of human life. Its exponential growth \cite{ros22} confronts IoT networks with tremendous challenges as connected devices generate massive and increasing amounts of data \cite{hajian22}. To address these challenges, it is crucial to combine IoT networks with emerging technologies, such as network softwarization \cite{shamsan22}.

Network softwarization is an emerging approach that transforms the network into an open ecosystem by decoupling software and hardware components. This transformation is expected to improve network scalability while reducing costs \cite{shamsan22}. Software Defined Networking (SDN) and Network Function Virtualization (NFV) are among the most important softwarization techniques \cite{amin21}. 

Machine Learning (ML), a subset of Artificial Intelligence (AI), is a promising way to bring intelligence to the network \cite{sellami20} and is considered an enabler for effective network softwarization for IoT. Using ML techniques, networks can learn from their past experiences, increasing their resilience to vulnerabilities while maximizing performance.

The combination of IoT, network softwarization, and ML, and the consideration of a distributed SDN controller design, allows to move towards a distributed intelligent network softwarization for IoT \cite{zormatimass23}, where these promising technologies combine forces to alleviate the challenges (e.g., energy consumption, scalability, etc.). This novel architecture is considered as a baseline to address many pressing concerns facing IoT networks, of which is routing \cite{samadi22}.

Although many works propose to optimize routing for softwarized networks, few of them consider the peculiarities of IoT, and to the best of our knowledge, no work has considered routing techniques based on Federated Deep Reinforcement Learning (FDRL), which is one of the most promising ML techniques enhanced by using a distributed intelligent network softwarization architecture. In this work, we propose a FDRL-based intelligent routing solution for softwarized IoT, where the network takes advantage of intelligence and distributed control design to extend the learning capabilities and achieve optimal dynamic routing decisions.

The remainder of the paper is organized as follows. In Section 2, we give a brief overview of related work. In Section 3, we present the most important background notions. In Section 4, we formalize the routing problem. Then, in Section 5, we present our routing optimization solution with a detailed description of the intelligent FDRL part. In Section 6, we present the experimental setup and simulation results. Finally, we conclude the paper in Section 7.

\begin{figure*}[htbp]
\centerline{\includegraphics[scale=0.36]{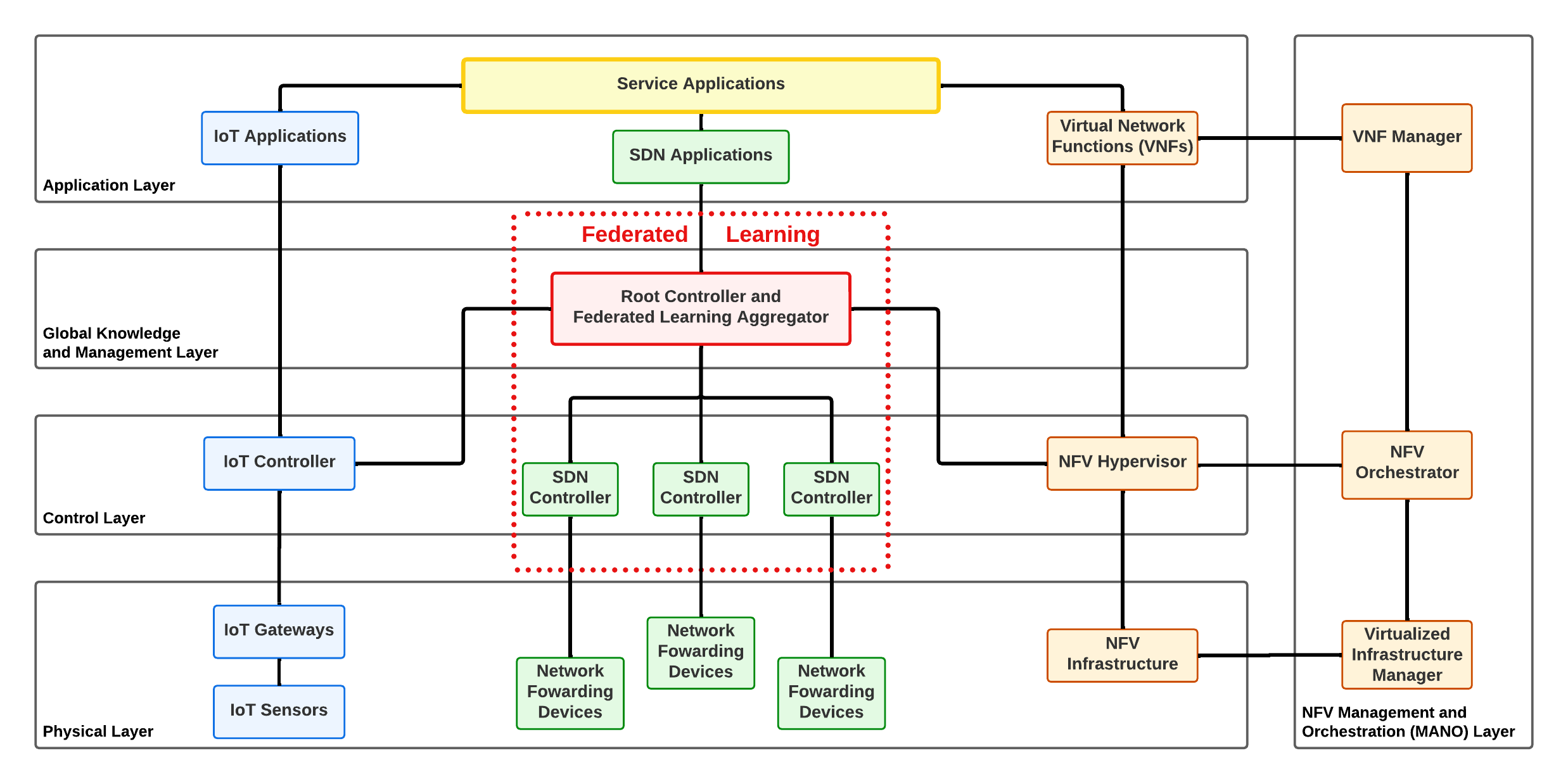}}
\caption{Distributed Intelligent Network Softwarization Architecture for IoT \cite{zormatimass23}}
\label{fig}
\end{figure*}

\section{Related work}
In IoT networks, the sheer number of heterogeneous IoT devices poses tremendous challenges, which opens many research areas to build dynamic IoT networks and overcome the constraints.

Many works have been done to alleviate the limitations of IoT by considering its combination with network softwarization (mainly, SDN and NFV), or ML, or ideally both, moving towards intelligent network softwarization for IoT. Although it is evident that academia is fully aware of the potential of SDN, NFV, ML, and IoT as key enabling technologies, the authors in \cite{zormatisoftcom23} note that the works exploiting the full potential of these technologies in combination are still sparse.

Routing is one of the most challenging networking applications in IoT, given the large number of interconnected nodes. Although considerable work has been done in this area \cite{hajian22} \cite{ouhab20} \cite{samadi22} \cite{shuker19}, to the best of our knowledge, no work has been done to consider distributed intelligent approaches (i.e., Federated Learning (FL)-based solutions) to optimize routing in constrained IoT environments. This can be explained by the fact that the conventional IoT network architecture does not make it easy to move towards such approaches. However, thanks to network softwarization (mainly, SDN with a distributed controller design), it is easier to move towards distributed intelligent routing for IoT. Therefore, our current work aims to fill this gap.

\section{Background}

In order to achieve distributed intelligent network softwarization for IoT, having as motive to satisfy the requirements in terms of Quality of Service (QoS) and energy efficiency, the authors in \cite{zormatimass23} propose the efficient general architecture shown in Figure \ref{fig}. It is built based on the reference architectures of IoT, SDN, NFV, and ML, thus obtaining an intelligent network softwarization for IoT.

With the exponential growth of IoT, relying on a single SDN controller raises concerns about its reliability \cite{xiao18}. Therefore, in the considered architecture, a distributed hierarchical SDN controller design is adopted, where multiple controllers coexist and are coordinated by a root controller. This enables distributed intelligent network softwarization for IoT, and such a design facilitates the implementation of distributed learning approaches such as FL.

As an alternative to traditional ML techniques, FL not only allows autonomous decision making when based on appropriate ML techniques, but also allows building global knowledge without sharing raw data among learning nodes, thus improving privacy and communication efficiency \cite{nguyen21}. In the architecture presented above (see Figure \ref{fig}), each SDN controller acts as a learning node, and the root SDN controller acts as an FL aggregator.

Routing, one of the most challenging aspects of large-scale IoT networks, is considered as a decision problem, and thus Reinforcement Learning (RL) is the most appropriate ML technique to obtain optimal routing decisions that are dynamically adapted \cite{amin21}. RL is designed to teach an agent to make decisions and take actions to maximize a long-term cumulative reward (based on feedback from the environment). The goal is to learn the optimal behavior (i.e., policy) to maximize the reward.

Given the complexity of IoT networks, which induces larger input state and action spaces, the combination of RL and Deep Neural Networks (DNNs) should be considered, resulting in Deep Reinforcement Learning (DRL). By exploiting the capabilities of DNNs, DRL addresses the limitations of traditional RL (e.g., low convergence rate to the optimal policy, inability to solve problems with high-dimensional input spaces, etc.).

Therefore, based on the presented general architecture for IoT networks, we propose to optimize routing using FDRL, the combination of FL and DRL that takes advantage of both techniques. The proposal benefits from the massive amount of available data to train a global model that ensures the most accurate routing decisions.

\section{Problem formulation}

Formally, we represent the network as a directed graph \(G(V,E)\), where \(V\) denotes the set of switches and routers, and \(E\) denotes the set of links between them, such that \(E=\{(i,j)\in V\times V, i\ne j\}\). For each link \((i,j) \in E\), the delay (i.e., latency), loss, and bandwidth metrics are represented by \(\{d_{ij},l_{ij},b_{ij}\}\in \mathbb{R}_{+} \), respectively. 

Given a source node \(\alpha\) with a flow \(f\) that has as destination node \(\beta\), our goal is to find a path that minimizes delay and loss while maximizing throughput, thus ensuring good QoS. Since we also need to improve the energy efficiency, we should reduce the number of hops and thus the energy consumption for the flow transmission. We refer to the number of hops of a given flow \(f\) from \(\alpha\) to \(\beta\) as \(h_{\alpha,\beta,f}\). This metric can vary because the route can be different from one flow to another for the same source-destination pair. 

To facilitate modeling, we adopt a discrete time model where time is divided into consecutive time slots \(t=1,2,...\), and the \(t\)-th flow arrives at the beginning of time slot \(t\). It is assumed, without loss of generality, that flow requests arrive sequentially. For a flow \(f\) arriving at time \(t\), we note \(U(t)\) the utility function of the \(t\)-th flow. The objective is then:

\begin{equation}
maximize \lim_{{T \to \infty}} \sum_{{t=1}}^{T} U(t)
\end{equation}

For this purpose, we propose to consider an FDRL agent composed of an FL aggregator (i.e., the root controller) and FDRL learning nodes (i.e., the SDN controllers).

\section{Proposed solution}

Here, we present our proposed routing optimization solution for softwarized IoT to meet the QoS and energy efficiency requirements that traditional routing mechanisms such as Shorest Path Routing (SPR) cannot meet as networks grow and become increasingly difficult to model and control. This solution takes full advantage of distributed intelligence by being based on FDRL.

\subsection{Solution overview}

Consider an IoT network where any node can communicate with any other node at any time. Each flow has specific requirements on a set of QoS metrics (e.g., latency, throughput, loss). Since we are evolving in a constrained environment, each hop implies an energy consumption (due to data transmission and processing at intermediate nodes). Therefore, from an energy efficiency point of view, it is better to reduce the number of hops. Our goal is to dynamically decide which route to take for each communication flow. The flow should be successfully forwarded from the source to the destination while satisfying the quality and energy requirements.

Since we are in a distributed softwarization architecture, each SDN controller is responsible for a part of the network. DRL is considered as the most appropriate technique for online decision problems, of which routing is one \cite{jalil2020} \cite{liu2021}. The architecture proposed in \cite{zormatimass23} allows us to evolve towards FDRL, which allows controllers to have stronger routing policies as they are trained on larger datasets without having to exchange raw data, which would have increased communication overhead and opened up potential security failures \cite{hwang22}. 

\begin{longtblr}[
  label = tb:comprt,
  caption = {Comparison of SPR, DRL-R, and FDRL-R Approaches},
]{
  width = \linewidth,
  colspec = {Q[237]Q[215]Q[229]Q[260]},
  hline{1} = {2-4}{},
  hline{2-8} = {-}{},
}
 & \textbf{SPR} & \textbf{DRL-R} & \textbf{FDRL-R}\\
\textbf{Complexity} & Low (e.g., Dijkstra) & High (DNN) & Extremely high\\
\textbf{Need for training} & No & Yes & Yes\\
\textbf{Real-time adaptability} & Static & Highly adaptive & Highly adaptive\\
\textbf{Multi-objective} & No & Yes & Yes\\
\textbf{Scalability} & Low (good for small networks) & Good (scales well with network size) & Excellent (scales across multiple domains)\\
\textbf{Privacy} & Not considered & Not considered & Designed for data privacy

\end{longtblr}

In Table \ref{tb:comprt}, we propose a comparison of SPR, DRL routing (DRL-R), and FDRL routing (FDRL-R). We note that although DRL-based routing methods require prior training, they guarantee real-time adaptability while ensuring scalable and multi-objective routing decisions (e.g., handling different flows with different QoS requirements). Much work has been done and proven the effectiveness of DRL routing, and FDRL routing will surely replace its counterparts once all challenges have been addressed, as it is the most complicated approach (due to the need to handle both FL and DRL complexities). Despite the complexity concerns associated with FDRL, as it relies on learning algorithms that require computational resources for training and decision making, the justification for considering FDRL lies in the long-lasting benefits and advances in network performance, as it achieves a favorable trade-off between the solution complexity and the benefits it provides.

FDRL leverages the large-scale nature of IoT networks to gather collective knowledge, improving routing decisions for better network performance and lifespan. Its power lies in its adaptability, leveraging ML to dynamically adjust routing strategies based on learned experience while satisfying key constraints. This approach excels in IoT constrained environments by exploiting shared intelligence to optimize routing, something traditional methods may lack due to their static nature and limited ability to adapt to evolving network conditions. Our current work is the foundation for having a fully functional routing solution in the near future, especially if research efforts are combined.

\begin{figure*}[htbp]
\centerline{\includegraphics[scale=0.325]{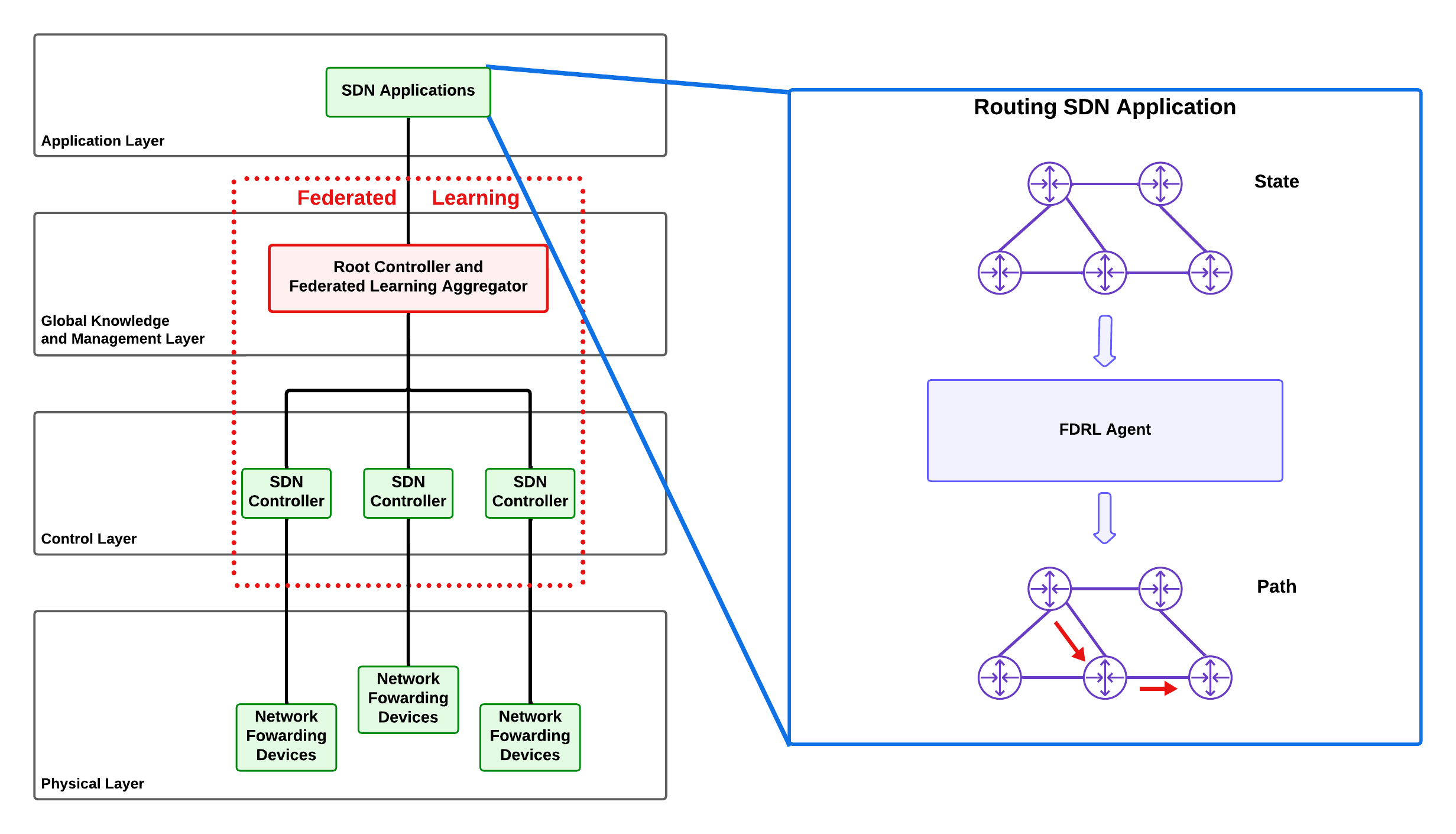}}
\caption{FDRL Routing for Softwarized IoT}
\label{fig:routarch}
\end{figure*}

In Figure \ref{fig:routarch}, we present an overview of the routing solution. It is consistent with the general IoT network softwarization architecture. For the sake of clarity, we keep only a subset of the SDN-related elements. We recall that routing is one of the multiple SDN applications. It periodically gathers network statistics to ensure that the selected path for a source-destination pair request remains consistent with the current network state.

\subsection{Description of the DRL part}
We now present the DRL mechanism that finds a path for a flow \(f\) from \(\alpha\) to \(\beta\) while optimizing the QoS and energy efficiency parameters as described earlier.

The state space is designed to capture the necessary information about the current state of the network. Therefore, we capture delay, loss, and throughput (all of which are QoS metrics). Hence, energy efficiency is evaluated based on the number of hops that make up the decided path. 

The action space consists of all edges (i.e., links) of the network. The action space is defined as \(A=\{a_{1},a_{2},...,a_{|E|}\}\), where each action corresponds to a link in the network \((i,j)\in E\). Actions are taken by the agent based on the observed state and in order to maximize the long-term cumulative reward.

The reward function is what directly affects the learning of the DRL algorithm. In our proposal, the reward function includes signals to deal with the optimization of the multiple metrics (i.e., maximizing the utility function), but also invalid actions and loops that may occur because the agent can decide any action at any time. Let \(p\in \mathbb{R}_{-}\) be a penalty value. It should be set in a way to have the best balance between exploration of new paths and exploitation of previous knowledge. The reward function is defined as:

\begin{equation}
r(t) =
\begin{cases}
  p & \text{if invalid} \\
  U(t) & \text{else}
\end{cases}
\end{equation}

The agent's goal is to accumulate the maximum positive reward in each episode in order to have the maximum long-term cumulative reward. The utility function is defined as follows:

\begin{equation}
U(t)=w_{1}x_{t}-w_{2}d_{t}-w_{3}l_{t}-w_{4}h_{t}
\end{equation}

\noindent where \(x_{t}\) is the throughput to maximize, \(d_{t}\) is the delay to minimize, \(l_{t}\) is the loss to minimize (i.e., maximize the Packet Delivery Ratio (PDR)), and \(h_{t}\) is the number of hops to minimize. All these metrics are weighted by \(w_{i}\in \mathbb{R}_{+}, i\in \{1,2,3,4\}\), which can be tuned to prioritize any metric during link selection.

We use the defined state space, action space, and reward function for DRL. To take an action, DRL uses DNN to approximate the reward that will result from the action. It outperforms RL in terms of handling larger state and action spaces. Our DNN takes the state space as input and returns a vector of action values as output.

We undertake a training process to enable our DRL agent to learn the optimal policy that maximizes the long-term cumulative reward. During this phase, our DRL algorithm iteratively updates the model parameters, represented as \(\theta\), in search of an improved policy. The end result of this training phase is the determination of the target model weights \(\theta^*\), which represent the optimal parameter configuration for our agent's policy.

\subsection{Design of the FDRL routing}

\begin{figure*}[htbp]
\centerline{\includegraphics[scale=0.5]{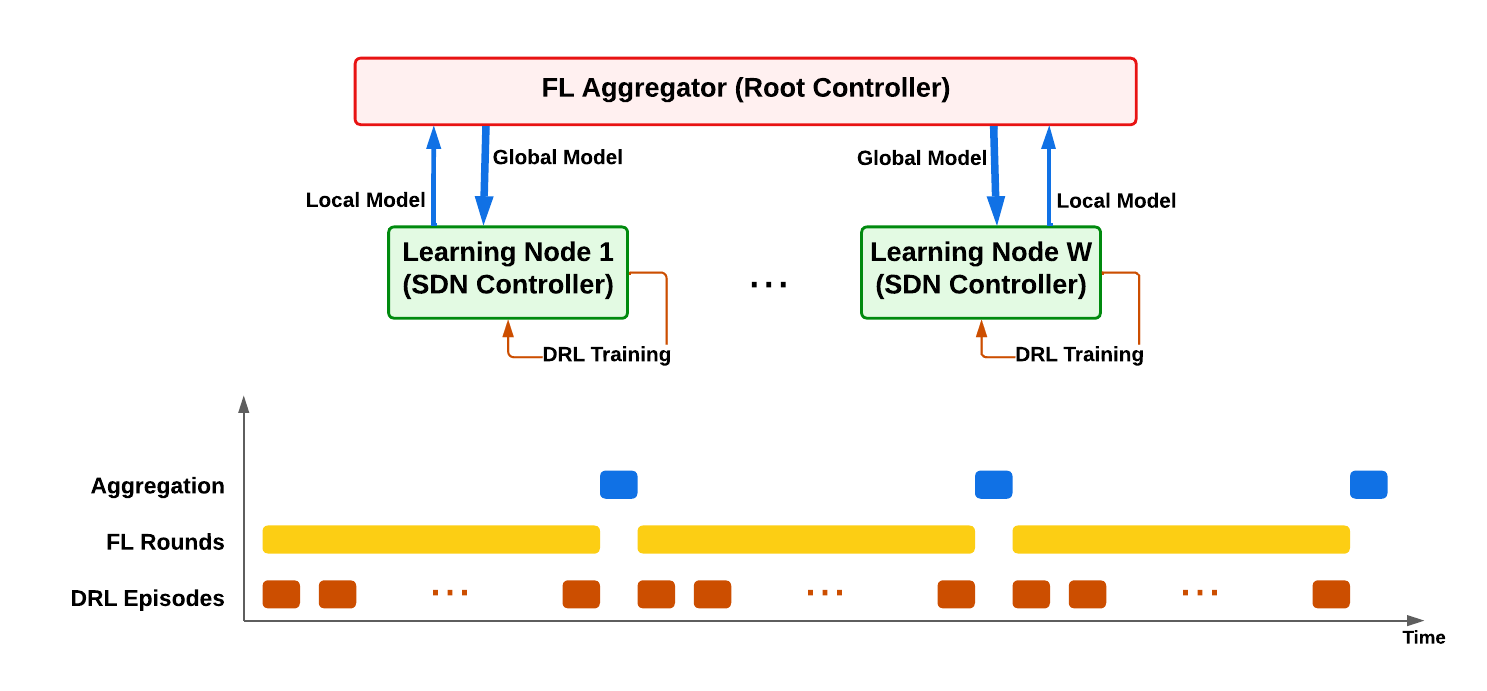}}
\caption{FDRL Training Phase}
\label{fig:fdrlb}
\end{figure*}

We build on the existing work on performing DRL routing in softwarized networking environments \cite{jalil2020} \cite{liu2021} \cite{tehrani2021}, and on the description of the DRL part presented earlier, to move towards a federated approach, namely FDRL routing. Figure \ref{fig:fdrlb} shows the behavior of our novel routing solution during the training phase.

As illustrated in the Gantt chart (cf. Figure \ref{fig:fdrlb}), each learning node, which corresponds to an SDN controller, trains its DRL agent locally, thus performing local weight updates \(\theta^e, e\in N_e\), for a number of episodes \(N_e\). After each FL round \(r\in R\), the local model weights are sent to the FL aggregator, which is the root controller, which performs aggregation and sends back the updated global model weights. At the end of the training, all nodes will have the final global model weights \(\theta^*\) and will be able to have an optimized routing policy that ensures QoS performance and energy efficiency.

The Algorithm 1 is proposed for FDRL aggregation. We use Federated Averaging (FedAvg) \cite{nilsson2018} for aggregation to obtain the global model weights \(\theta^r\). Initial weights \(\theta^0\) are sent by the aggregator, and for each round \(r\), local model weights are collected from the learning nodes, aggregated, and sent back as updated global model weights, and this is repeated until the end of the training. The final global model weights \(\theta^*\) are sent to all learning nodes.

\noindent\includegraphics[width=\columnwidth]{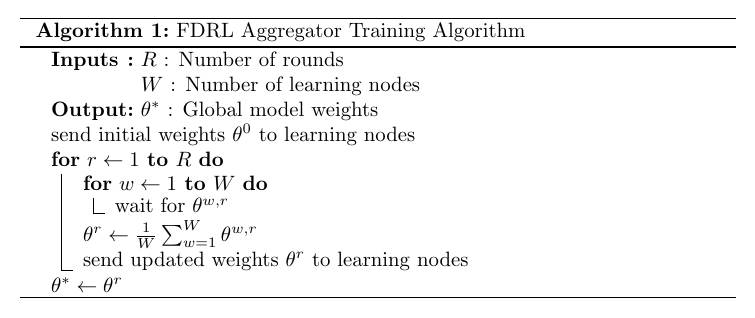}

We also propose Algorithm 2 for FDRL learning nodes training. Local DRL agents will perform \(N_e\) training episodes, each with a duration \(T\). Given the current state \(s_t\), the agent will decide and take an action \(a_t\) to get to state \(s_{t+1}\) and get immediate reward \(r_{t+1}\). When the time comes for aggregation, the local model weights are sent to the aggregator, which sends back the updated global model weights \(\theta^g\). At the end of the training process, each learning node will have the final global model weights \(\theta^*\). Note that the exploration parameter \(\epsilon\) impacts the balance between exploring new possibilities and exploiting the current best-known actions.

\section{Simulation results}

In this section, we demonstrate the effectiveness of our solution through simulation and performance evaluation. Considering the complexity of the proposal and the fact that it contains many novelties, we test each novelty (i.e., sub-solution) separately. This allows us to easily interpret the results by being able to identify the cause behind each effect. After introducing the simulation environment and the performance metrics to be observed, we present the experimental results and discuss the impact of distributed SDN control and intelligent routing on softwarized IoT networks. We also justify the overall effectiveness of our solution.

\noindent\includegraphics[width=\columnwidth]{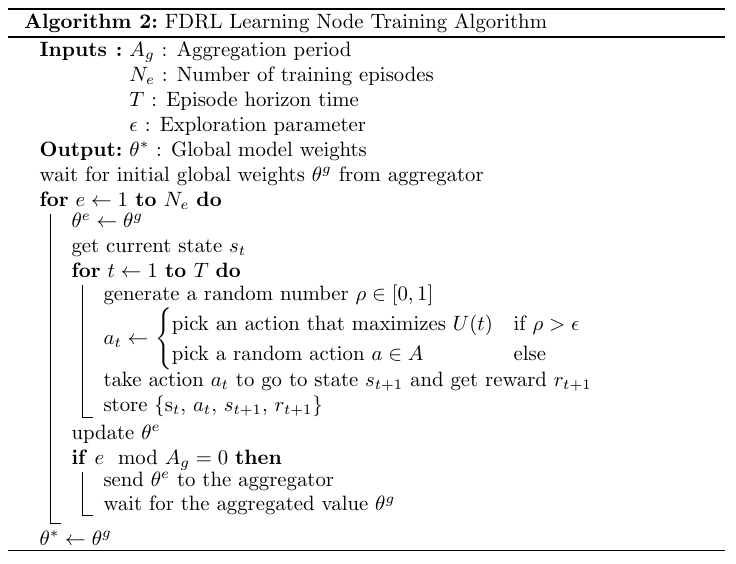}

\subsection{Simulation environment}

We now introduce the hardware and software simulation environment. In particular, we focus on explaining the choice of the network simulation tool and the SDN controller. As for the hardware environment, the simulations are performed on a computer equipped with an AMD Ryzen 7 4000 series processor supported by 16 GB of memory. The computer runs on Ubuntu 22.0.4, a Linux distribution. We use Python as the main programming and scripting language.

\newpage

In order to carry out the experiments, we proceed with Mininet, which is well positioned compared to its counterparts. It is a network emulator that facilitates the creation of network topologies for research purposes. It allows the emulation of networks with virtual hosts, switches, and controllers, all running on real Linux kernels. It makes it easy to prototype and experiment various scenarios. We recall that network emulation provides better and more accurate results than simulation because it replicates real-world behavior by running the network software on virtualized hosts \cite{patel19}.

For the SDN controller, we opt for the Ryu SDN controller, which is one of the most powerful controllers, and the choice is strongly driven by the fact that there is a strong complementarity between Mininet and Ryu. Ryu is an open source SDN controller that offers extensive programmability, which has made it a popular choice for network research.

\subsection{Results and discussion}

Before introducing the results, we present the key metrics used in the evaluation. Hence, we present the delay, throughput, and loss ratio metrics, summarized in Table \ref{tb:metrics}. Further evaluation may be based on other QoS or energy-related performance indicators.

\begin{longtblr}[
  label = tb:metrics,
  caption = {Summary of Considered Metrics},
]{
  width = \linewidth,
  colspec = {Q[198]Q[265]Q[100]Q[150]},
  hline{1} = {-}{},
  hline{2-7} = {-}{},
}
\textbf{Metric} & \textbf{Description} & \textbf{Unit} & \textbf{Ideal value}\\
\textbf{Delay} & Measures data travel time from source to destination. & ms & The lowest\\
\textbf{Throughput} & Measures data transmission rate in a network. & Mbps & The highest\\
\textbf{Loss ratio} & Represents the proportion of lost data packets during transmission. & \% & The lowest

\end{longtblr}

After setting up the environment, we run simulations to determine the effectiveness of our proposal. As presented, we proceed in two main steps. First, we evaluate the effect of distributed SDN control on the performance of IoT networks. Then, we evaluate the effect of intelligent routing on the performance of softwarized IoT networks.

\subsubsection{Effect of distributed control}~\\
Here, we perform simulations to infer the impact of a distributed SDN control architecture on network performance. We measure the delay and throughput metrics in the case of centralized control (i.e., one SDN controller) and in the case of distributed control (i.e., two or more SDN controllers). 

\begin{figure}[htbp]
\centerline{\includegraphics[scale=0.265]{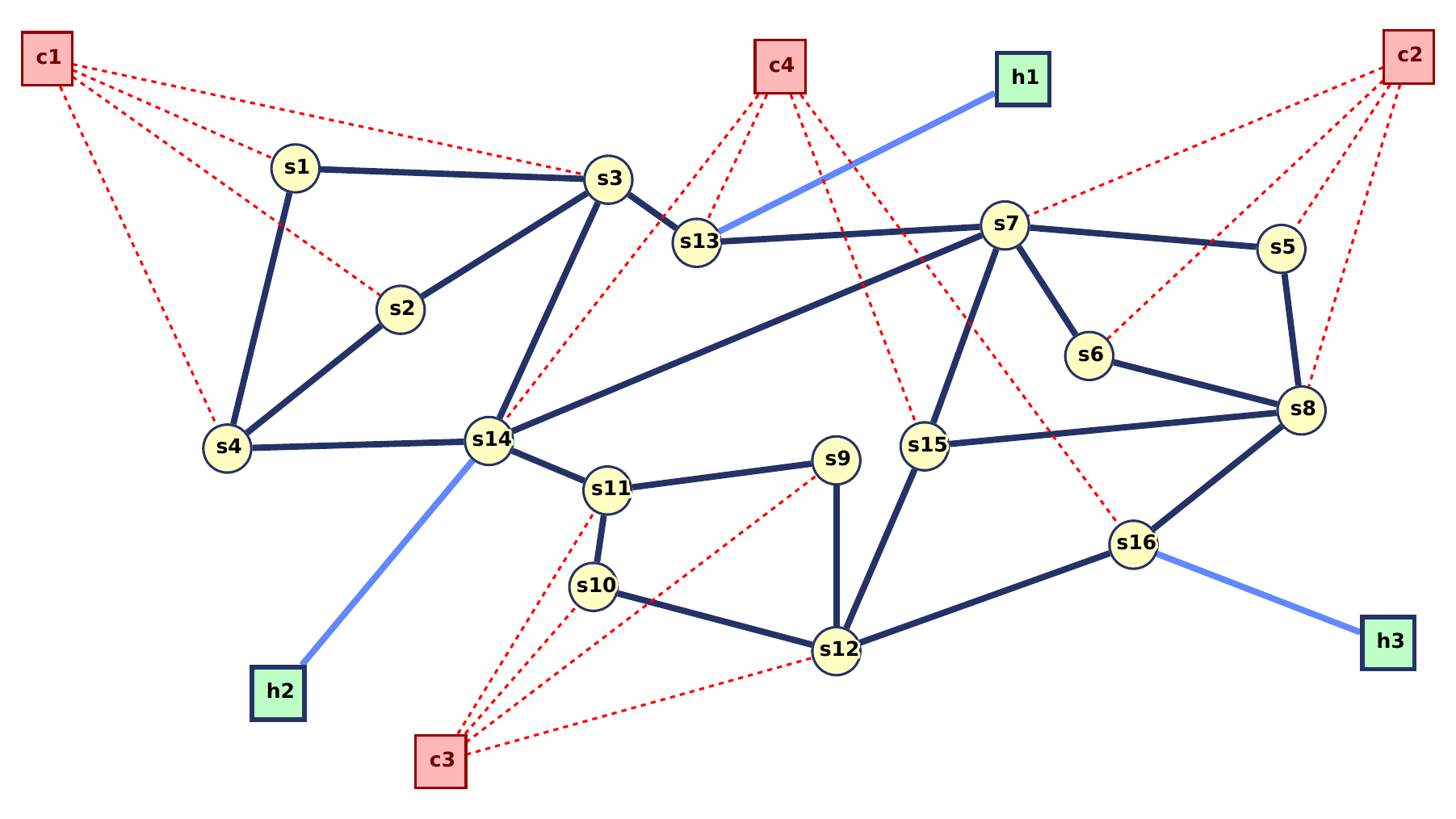}}
\caption{Adopted Topology to Evaluate Distributed Control}
\label{fig:topo1}
\end{figure}

We take the topology illustrated in Figure \ref{fig:topo1}. It consists of 16 forwarding devices and 3 hosts \(\{h1,h2,h3\}\). We propose 3 test cases: 4 SDN controllers \(\{c1,c2,c3,c4\}\), each controlling a group of devices (as in Figure \ref{fig:topo1}), 2 SDN controllers where \(c1'\) controls the devices of \(\{c1,c2\}\) and where \(c2'\) controls the devices of \(\{c3,c4\}\), and 1 centralized SDN controller (i.e., the common SDN controller design).

We consider SDN controllers that use a basic routing strategy (i.e., SPR). We instantiate the Ryu controllers, run the Mininet script that creates and connects the hosts and switches, and connects the switches to the remote controllers. Using \textbf{\textit{ping}}, a widely used network diagnostic and testing tool, we take 250 Round-Trip Time (RTT, or latency) measurements with 64 bytes of data to evaluate the delay when \(h1\) pings \(h2\), when \(h1\) pings \(h3\), and when \(h2\) pings \(h3\).

\begin{figure*}[htbp]
\centerline{\includegraphics[scale=0.41]{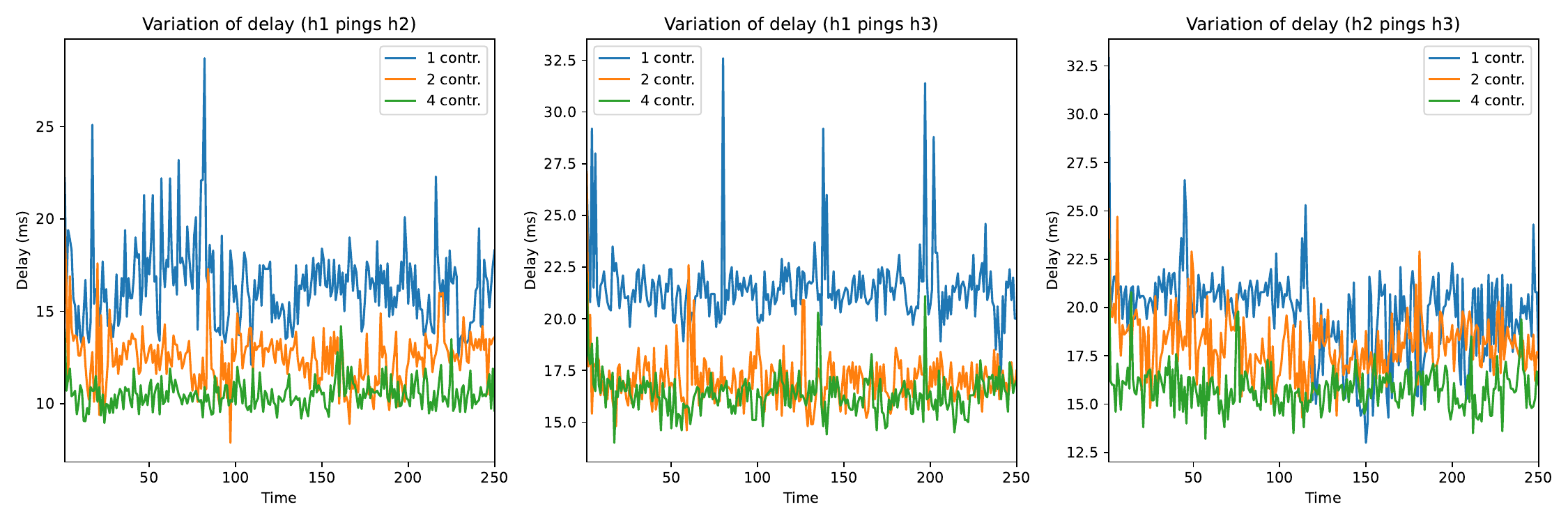}}
\caption{Variation of Delay with Time}
\label{fig:gr1}
\end{figure*}

We perform the measurements for three cases: with 1, 2, and 4 SDN controllers leading the network forwarding devices. The results are shown in Figure \ref{fig:gr1}. For each source-destination pair, we plot the delay values for the three cases. 

We observe that for the three source-destination pairs, as we increase the number of SDN controllers, moving towards a distributed control design, the delay decreases for almost all the measures. Delay varies from one measure to another because it depends on the network state, but also due to the instability of the emulation environment (since the measures are performed in a virtual environment).

\begin{figure*}[htbp]
\centerline{\includegraphics[scale=0.22]{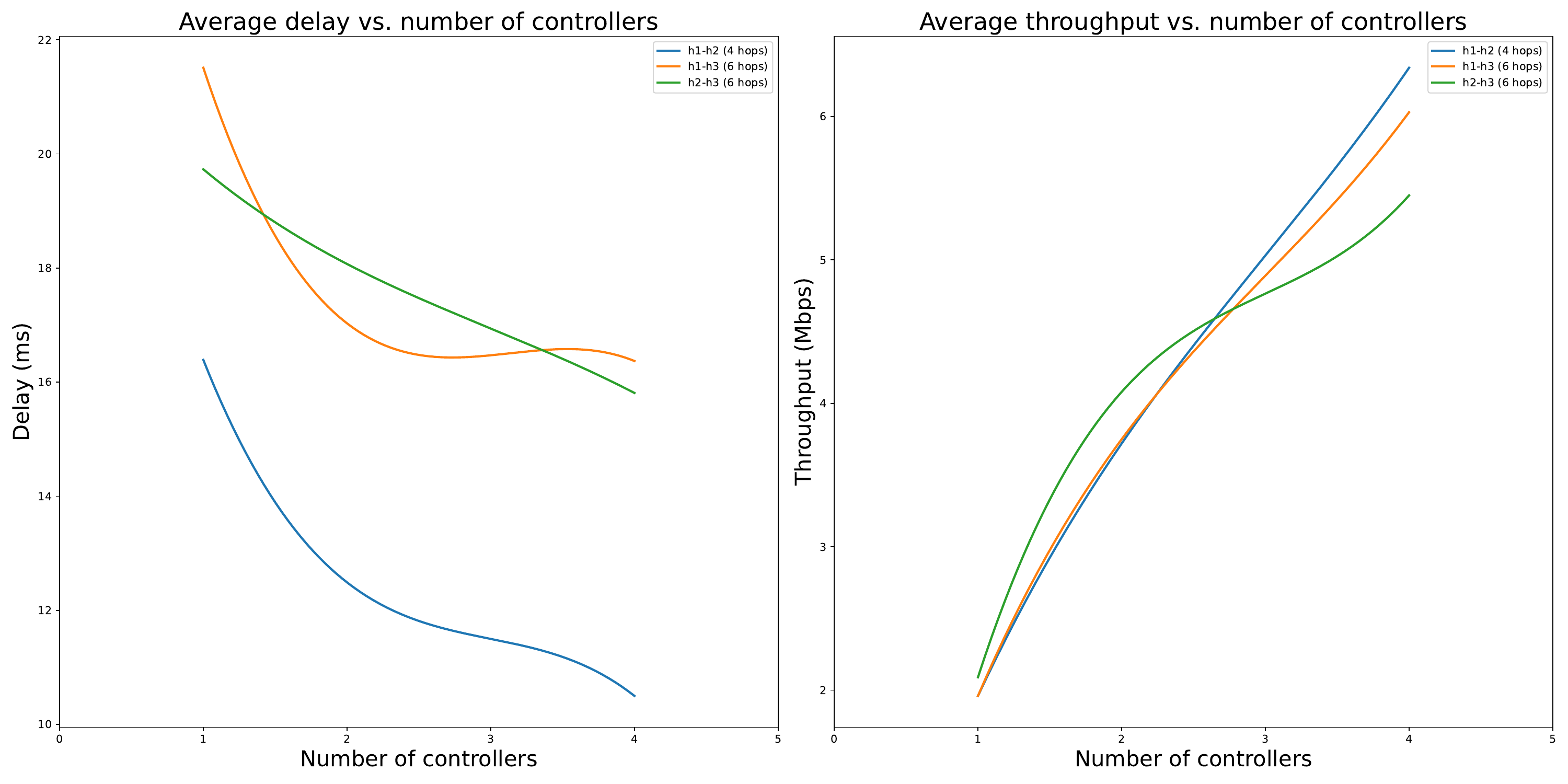}}
\caption{Average Delay and Throughput}
\label{fig:gr2}
\end{figure*}

Before drawing conclusions, we also evaluate the average throughput between the same source-destination pairs and for the three test cases (i.e., 1, 2, or 4 controllers). For this purpose, we use the \textit{\textbf{iperf3}} tool. It is a powerful network performance testing tool designed to measure the throughput, bandwidth, and quality of network connections. We perform a 100 second duration test, for three times. To test the throughput between \(h1\) and \(\{h2, h3\}\), the host \(h1\) acts as \textit{\textbf{iperf3}} server and the others as clients. Similarly, \(h2\) acts as a server and \(h3\) as a client to measure the throughput between \(h2\) and \(h3\). In Figure \ref{fig:gr2}, we plot the average delay and average throughput versus the number of controllers (1, 2, or 4).

We notice that the average delay observed between \(h1\) and \(h2\) is always lower than the other source-destination pairs. This is justified when we calculate the number of hops of the shortest path between the hosts using the Dijkstra algorithm: \(h1\) and \(h2\) are separated by 4 hops, while the other two pairs are separated by 6 hops. We do not observe this difference in the average throughput graph, which can be explained by the fact that there is no loss in the network and that all links of the topology have the same bandwidth setup.\\
~\\
~

As the number of SDN controllers increases, moving from a centralized to a distributed controller design, the delay decreases while the throughput increases, meaning that the overall network performance increases. Based on that, we conclude that moving towards distributed control has a great potential to increase the overall network performance as shown by the evaluated quality metrics. One of the main reasons for this is that the distributed design reduces the load on the controllers, thus speeding up the processing time, especially for large-scale IoT networks.\\

~

\subsubsection{Effect of intelligent routing}~\\
Now, we perform the necessary simulations to evaluate the outcome of having intelligence in the routing process in softwarized IoT networks. For this purpose, we consider the DRL-based routing method proposed by authors in \cite{liu2021}. We note that our context here deviates from the cited authors. We aim herein to demonstarte the effectiveness of intelligent DRL-based routing, to be applied in softwarized IoT.

We evaluate this intelligent routing approach in two common network topologies found in IoT scenarios: Abilene and GEANT. The reason for having two topologies is to challenge the DRL agent, since GEANT has more nodes and links than Abilene, and thus larger input state and action spaces. In the conducted evaluation, one controller is responsible of all the network. A 10\% random packet loss is added to the network. DRL routing is evaluated in terms of delay, throughput, and loss ratio.

\begin{figure*}[htbp]
\centerline{\includegraphics[scale=0.41]{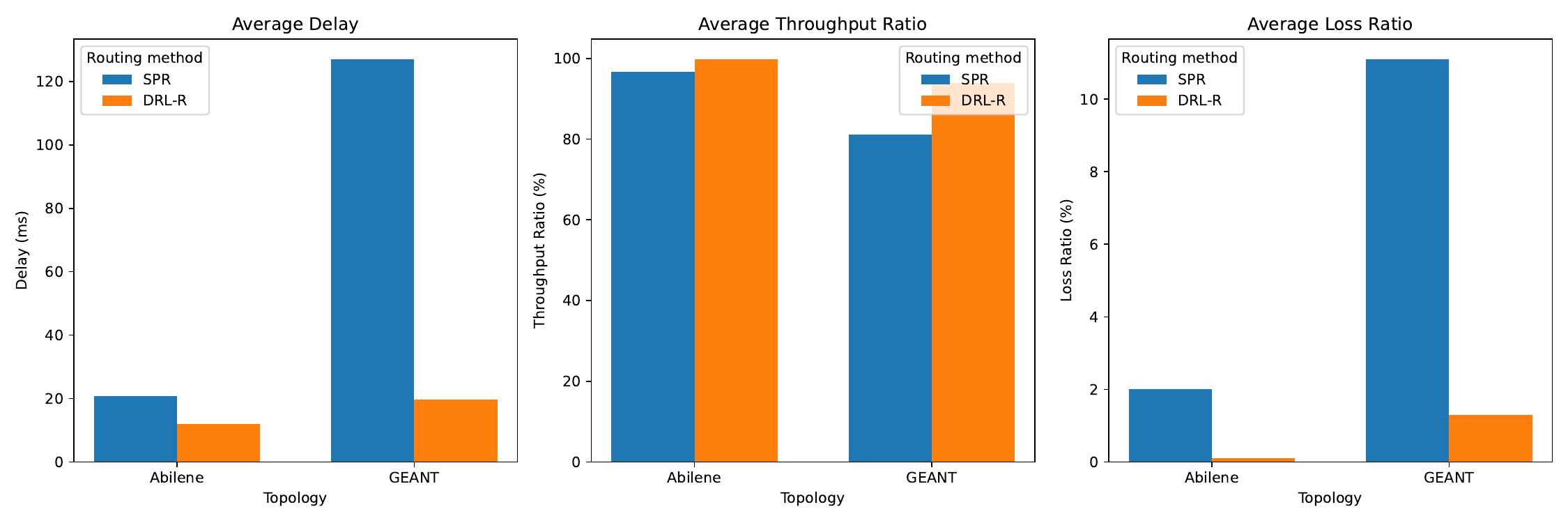}}
\caption{Intelligent Routing Evaluation Results}
\label{fig:gr3}
\end{figure*}

Results are presented in Figure \ref{fig:gr3}. We present the average latency, the average throughput ratio (calculated according to bandwidth), and the average loss ratio. For all scenarios and topologies, and for all three metrics, intelligent DRL-based routing outperforms conventional SPR. We also note that DRL routing still outperforms SPR in the GEANT topology, although it is more challenging because it is larger and causes a much larger input state and action spaces. We recall that this problem is one of the main reasons for moving from RL to DRL, since DNN allows us to handle larger state and action spaces.

Based on these results, we conclude on the real effectiveness of having intelligence in the network, especially for routing optimization, as it outperforms its counterparts in terms of improving performance metrics (lower delay, higher throughput ratio, and lower loss ratio).

Considering the results of our two main experiments and the conclusions drawn, we finally conclude that the presence of distributed control and intelligence in the network greatly improves the routing, and thus improves the overall performance of the network. Since there are no conflicts or redundancies between these two novelties, we strongly support the effectiveness of the overall solution.

\section{Conclusion}
Given the exponential growth rate of IoT networks, which are becoming increasingly complex to manage, it is now challenging for conventional routing approaches to achieve high performance and satisfy all requirements (e.g., QoS, energy efficiency). In this work, we proposed a novel FDRL-based routing approach to optimize performance based on a distributed intelligent network softwarization architecture for constrained IoT, and simulation results confirmed the effectiveness of our proposal.

As future work, we plan to fully implement the distributed intelligent network softwarization architecture to take advantage of NFV to dynamically enable and disable the routing network functions, thus improving the proposed routing approach and maximizing the overall network performance and lifetime. We also plan to consider the security aspect, as it is a very important concern that must be fully satisfied to achieve real-world applications of IoT.